\documentstyle[twoside,espcrc2-mod]{article}
\newcommand{\B}{{\em BeppoSAX\/}}
\newcommand{\U}{4U1626--67}
\newcommand{\chidof}{\mbox{$\chi^2_\nu$}}
\newcommand{\pmt}[2]{_{-#1}^{+#2}}
\newcommand{\tablenotemark}[1]{\rlap{$^{\rm #1}$}}
\title{\B\ observations of the X--ray binary pulsar \U}

\author{M. Orlandini\address{%
  Istituto Tecnologie e Studio Radiazioni Extraterrestri (TeSRE), C.N.R., \\
  \ via Gobetti 101, 40129 Bologna, Italy},
D. Dal~Fiume$^a$, F. Frontera$^a$\thanks{Also Physics Dept., Ferrara
  University},
S. Del~Sordo\address{%
  Istituto Fisica Cosmica e Applicazioni all'Informatica (IFCAI), C.N.R., \\
  \ via La Malfa 153, 90146 Palermo, Italy},
S. Piraino$^b$, A. Santangelo$^b$, A. Segreto$^b$,
T. Oosterbroek\address{%
  Astrophysics Division, Space Science Department of ESA, ESTEC, \\
  \ Keplerlaan 1, 2200 AG Noordwijk, The Netherlands} and
 A.N. Parmar$^c$}

\begin{document}

\begin{abstract}
We report on observations of the low-mass X--ray binary \U\ performed during
the \B\ Science Verification Phase. We present the broad-band 0.1--100 keV
pulse averaged spectrum, that is well fit by a two-component function: a
$0.27\pm 0.02$ keV blackbody and an absorbed power law with a photon index of
$0.89\pm 0.02$. A very deep and narrow absorption feature at 38 keV,
attributable to electron cyclotron resonance, is clearly visible in the
broad-band spectrum. It corresponds to a neutron star magnetic field
strength of $3.3\times 10^{12}$ G. The \U\ pulse profiles show a dramatic
dependance on energy: the transition between the low energy ($E<10$ keV) {\em
bi-horned\/} shape to the high-energy ($E>10$ keV) sinusoidal profile is
clearly visible in our data. The modulation index shows a monotonic increase
with energy.
\end{abstract}

\maketitle

\section{INTRODUCTION}

The low-mass X--ray binary system \U\ is formed by a neutron star pulsating at
about 7.7~s \cite{667} and orbiting the faint blue star KZ~TrA \cite{660}. This
system shows the lowest mass function ever observed \cite{610}, putting severe
constraints on the nature of the optical companion: the most likely is a
0.02--0.06 $M_\odot$ degenerate He or CO dwarf \cite{34}. The determination of
the \U\ orbital period have always been problematic: attempts to find Doppler
delays in the X--ray pulse arrival times have always been unsuccessful, but
photometric studies of the optical companion revealed the presence of two
periods: one at the X--ray pulsation frequency \cite{659} and another
down-shifted by about 0.4 mHz with respect to the first \cite{594}. Assuming
that the optical emission is due to the reprocessing of the X--ray emission on
the surface of the companion, a binary period of about 2500~s and a projected
semi-major axis of 0.4 lt-s are inferred \cite{1618}.

\section{OBSERVATIONS}

The \B\ satellite includes two Wide Field Cameras sensitive in the 2--30 keV
range \cite{1534}, and four Narrow Field Instruments (NFIs) sensitive in
0.1--10 keV (LECS \cite{1531}), 1--10 keV (MECS \cite{1532}), 3--180 keV
(HPGSPC \cite{1533}), and 15--300 keV (PDS \cite{1386}). It is a program of the
Italian Space Agency (ASI),  with participation of the Netherlands Agency for
Aerospace Programs (NIVR).

During the Science Verification Phase a series of well known X--ray sources
were observed in order to check the capabilities and performances of the
instruments on-board \B. \U\ is one of these sources and it was observed from
August 9 00:10:35 to August 11 00:00:05 UTC \cite{1622}. Data were telemetred
in direct modes, which provide complete information on time, energy and, if
available, position for each photon.

\section{SPECTRAL ANALYSIS}

\begin{table*}
\setlength{\tabcolsep}{1.5pc}
\newlength{\digitwidth} \settowidth{\digitwidth}{\rm 0}
\catcode`?=\active \def?{\kern\digitwidth}
\caption[]{Best-fit spectral parameters$^a$}
\label{tab:fit}
\begin{tabular}{lllll}
\hline
\multicolumn{2}{l}{Parameter} & \multicolumn{3}{c}{Value} \\
 \cline{3-5} & & No Line & Gaussian & Lorenzian \\ \hline
N$_{\rm H}$     & ($10^{21}$ cm$^{-2}$)      & $1.2\pm 0.2$           & $1.1\pm 0.2$           & $1.0\pm 0.2$\\
$kT$            & (keV)                      & $0.26\pm 0.02$         & $0.27\pm 0.02$         & $0.28\pm 0.02$ \\
$r_{\rm bb}$    & ($\times \rm d_{\rm kpc}$ km) & $1.7\pmt{0.2}{0.4}$ & $1.6\pmt{0.2}{0.3}$    & $1.5\pm 0.2$ \\
$\alpha$        &                            & $0.90\pm 0.02$         & $0.89\pm 0.02$         & $0.85\pm 0.02$ \\
E$_c$           & (keV)                      & $21.3\pmt{0.3}{0.4}$   & $19.8\pmt{0.5}{0.3}$   & $29\pmt{2}{3}$ \\
E$_f$           & (keV)                      & $7.8\pmt{0.4}{0.3}$    & $10.5\pm 0.8$          & $9\pmt{1}{2}$ \\
\tablenotemark{b}\ \ I$_{\rm pow}$ &         & $1.11\pm 0.03$         & $1.09\pm 0.03$         & $1.04\pm 0.04$ \\
E$_{\rm Ne}$    & (keV)                      & 1.05 \tablenotemark{c} & 1.05 \tablenotemark{c} & 1.05 \tablenotemark{c} \\
$\sigma_{\rm Ne}$ & (keV)                    & 0.04 \tablenotemark{c} & 0.04 \tablenotemark{c} & 0.04 \tablenotemark{a} \\
EW$_{\rm Ne}$   & (keV)                      & 48   \tablenotemark{c} & 48   \tablenotemark{c} & 48 \tablenotemark{c} \\
E$_{\rm CRF}$   & (keV)                      &                        & $37\pm 1$              & $33\pm 1$ \\
$\sigma_{\rm CRF}$ & (keV)                   &                        & $3\pm 1$               & $11\pm 2$ \\
EW$_{\rm CRF}$  & (keV)                      &                        & $14\pm 3$              &         \\
$\tau_{\rm CRF}$&                            &                        &                        & $1.5\pmt{0.3}{0.4}$ \\
\chidof         &                            & 1.482 (348)            & 1.258 (345)            & 1.122 (345) \\ \hline
\multicolumn{5}{l}{$^a$ Uncertainties at the 90\% confidence level for a single parameter.} \\
\multicolumn{5}{l}{$^b$ Flux is in units of $10^{-2}$~photons~cm$^{-2}$~s$^{-1}$ at 1 keV.} \\
\multicolumn{5}{l}{$^c$ The Ne line complex parameters are taken from \cite{1612}.}
\end{tabular}
\end{table*}

The pulse-phase averaged spectrum of \U\ has been described in terms of a two
component function: a blackbody with a temperature $kT\sim 0.6$ keV and an
absorbed power law \cite{658}. A high energy cutoff at $\sim 20$ keV is also
necessary to describe the high energy data \cite{617}. 

This spectral function describes very well our data, both the single \B\ NFIs
spectra, and the broad-band spectrum. We used the standard X--ray pulsar cutoff
of the form $\exp[(E_c - E)/E_f]$, where $E_c$ and $E_f$ are the cutoff and
folding energy, respectively. The smoother Fermi-Dirac cutoff did not
adequately describe the high energy tail. After the inclusion of the Ne line
complex at 1 keV \cite{1612} we obtained a best fit with a reduced \chidof\ of
1.482 for 348 degrees of freedom (dof). From the analysis of the residuals we
were led to add a cyclotron resonance feature (CRF) at $\sim 35$ keV. Both a
Lorenzian or a Gaussian in absorption model improved the fit, yielding \chidof
s of 1.122 and 1.258 for 345 dof, respectively. An F-test shows that the
improvement is significant at 99.99\%. The fit results are summarized in
Table~\ref{tab:fit} \cite{1622}. The search for a possible CRF at $\sim 19$
keV, as suggested by Pravdo \cite{617}, was unsuccessful.

The two CRF models are equivalent: the probability of chance improvement from
the Lorenzian to the Gaussian model is 30\%. However, we prefer the Gaussian
description of the CRF, because the cutoff energy in the Lorenzian fit is too
close to the resonance energy. As it is evident from Table~\ref{tab:fit}, the
Lorenzian model approximates the fall-off of the spectrum by increasing the
cutoff energy, and shifting the cyclotron energy at values lower than those
obtained by an absorption Gaussian. This is a known effect already observed for
the CRF in Her X-1 \cite{1583} and Vela X--1 \cite{1581}.

Our data do not show the presence of an Iron K-shell line in 6.4--6.9 keV: the
upper limit on its equivalent width is 21 eV, slightly more stringent than the
33 eV value obtained by ASCA \cite{393}.

The total 0.1--100 keV X--ray luminosity is $6.6\times 10^{34}$
erg~s$^{-1}$~d$^2_{\rm kpc}$. The fluxes in the bands 0.5--10 and 10--100 keV
are $1.5\times 10^{-10}$ and  $3.9\times 10^{-10}$ ergs~cm$^{-2}$~s$^{-1}$,
respectively.

\section{TIMING ANALYSIS}

\begin{figure*}
\vspace{16cm}
\includegraphics{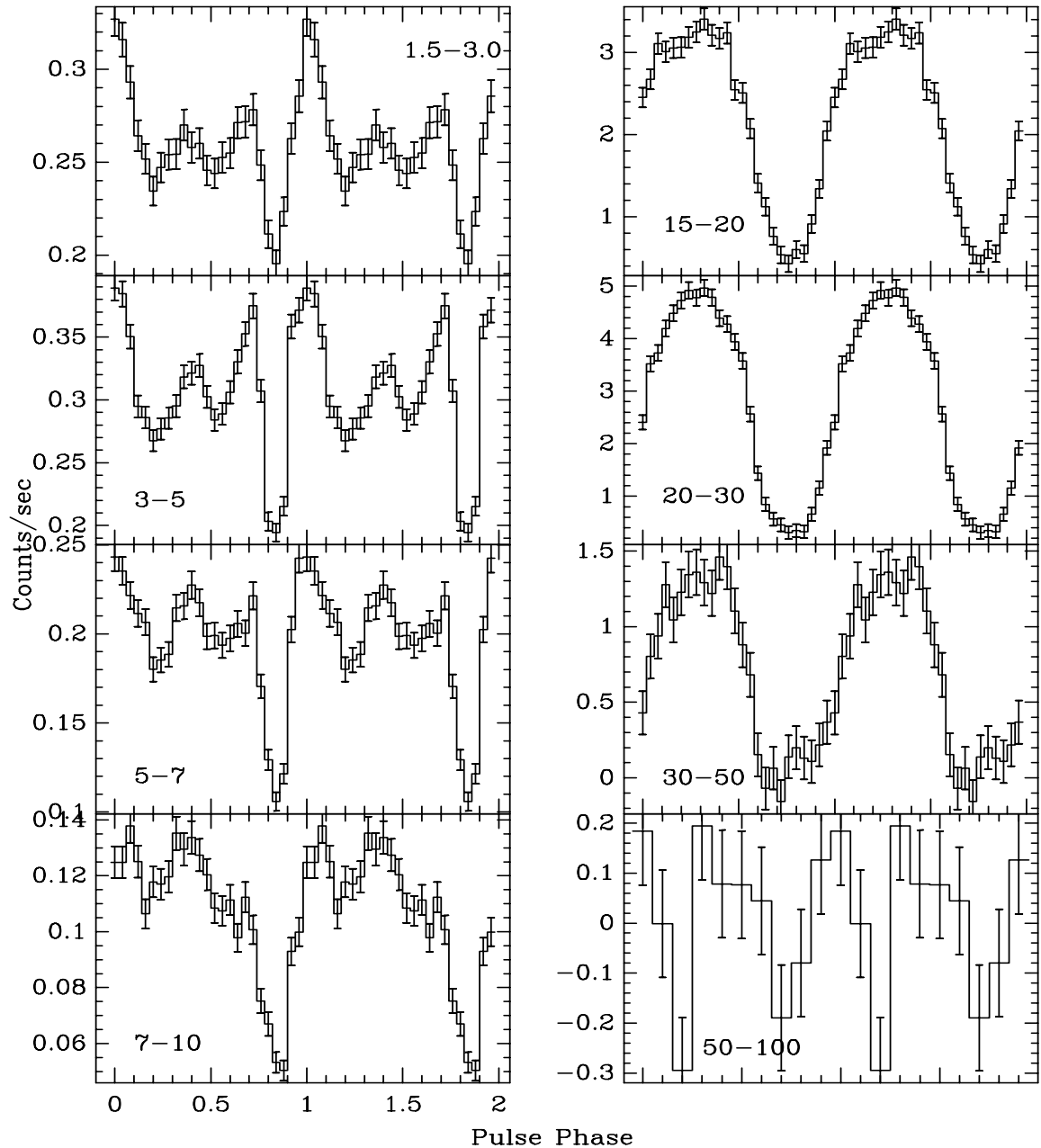}
\caption[]{\U\ pulse profiles as a function of energy observed by the MECS and
PDS instruments aboard \B. Note the transition between the {\em bi-horned\/}
shape to the almost sinusoidal form, attained with the increase of the
interpulse between the two {\em horns\/} as the energy increases.}
\label{fig:puls}
\end{figure*}

\begin{figure}
\vspace{6cm}
\includegraphics{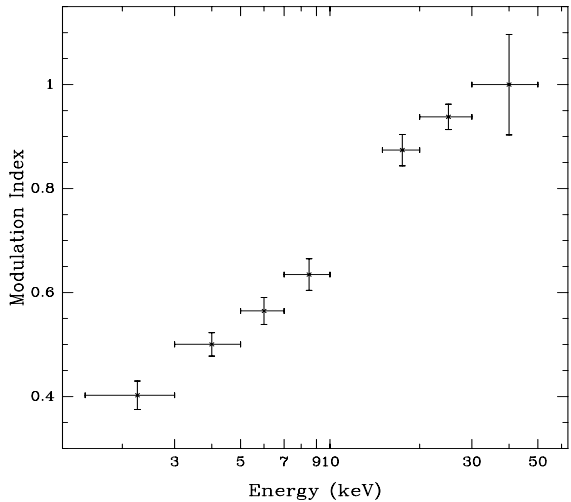}
\caption[]{The modulation index as a function of energy computed from the \U\
pulse profiles shown in Fig.~\ref{fig:puls}.}
\label{fig:mi}
\end{figure}

The pulse period referred to the Solar System barycenter is $7.66790 \pm
0.00005$~s, in agreement with BATSE results \cite{1591}. We folded the \B\
MECS and PDS data in different energy bands with this period, and the results
are plotted in Fig.~\ref{fig:puls}. We confirm the strong energy dependance of
the \U\ pulse profile. The 2--10 keV pulse shape presents the characteristic
{\em bi-horned\/} form, but the interpulse is not as flat as previously
observed: it shows the presence of a small peak. As the energy increases, this
third peak increases until it completely fills the interpulse, and the pulse
profile becomes sinusoidal. This is a clear indication of the anisotropy in the
radiative transfer in the strong neutron star magnetic field \cite{657}.

We also computed the so-called {\em modulation index\/}, defined $\Phi(E) = 1 -
I_{\rm min}(E)/I_{\rm max}(E)$ where $I_{\rm max}$ is the maximum intensity in
the pulse profile and $I_{\rm min}$ is the minimum, at the given energy $E$.
This index gives information on the dependence on  energy of the process which
determines the modulation \cite{1286}. The advantage of using the modulation
index instead of the pulse fraction is that the former does not require the
determination of the average emission. The result is shown in
Fig.~\ref{fig:mi}, where we can see a monotonic increase of the index with
energy.

\end{document}